\documentclass[preprint,12pt]{elsarticle}

\usepackage{amssymb}
\usepackage{graphicx}
\usepackage{subfigure}
\usepackage{tabularx}
\usepackage{adjustbox}
\usepackage{multirow}
\usepackage{amsmath,amsfonts}
\usepackage{float}

\journal{Intelligent Systems with Applications}

\begin{document}

\begin{frontmatter}


\tnotetext[label1]{This research is funded by the "Grants for Development of Education Innovation, Ratchadaphiseksomphot Endowment Fund, Chulalongkorn University", the "Newton Mobility Grants NMG\textbackslash R1\textbackslash 180310", and the "scholarship from the Graduate School, and Chulalongkorn University to commemorate
the 72nd anniversary of his Majesty King Bhumibol Aduladej".}
\cortext[cor1]{Corresponding author}

\title{Red Blood Cell Segmentation with Overlapping Cell Separation and Classification from an Imbalanced Dataset}


\author[inst1]{Korranat Naruenatthanaset}

\author[inst4]{Duangdao Palasuwan}
\author[inst5]{Nantheera Anantrasirichai}
\author[inst4]{Attakorn Palasuwan}
\author[inst1,inst3]{Thanarat H. Chalidabhongse \corref{cor1}}\ead{thanarat.c@chula.ac.th}

\address[inst1]{Department of Computer Engineering,Chulalongkorn University,Bangkok,Thailand}

\address[inst4]{Oxidation in Red Cell Disorders Research Unit,
            Chulalongkorn University, 
            Bangkok,
            Thailand}
\address[inst5]{Visual Information Laboratory,
            University of Bristol, 
            Bristol,
            BS8 1UB,
            UK}
\address[inst3]{Applied Digital Technology in Medicine (ATM) Research Group,
            Chulalongkorn University, 
            Bangkok,
            Thailand}

\begin{abstract}
Automated red blood cell (RBC) classification of blood smear images can reduce time and cost of the RBC analysis process. Two main challenges of this task are: i) overlapping cells leading to incorrect prediction, so cell separation is required; and ii) imbalanced dataset, where the number of normal class are significantly larger than those of rare disease classes. In our case, the dataset contains 12 RBC classes and the highest imbalance ratio is 34.538, hence more challenging than many other applications.
This paper presents a new framework of RBC identification in the blood smear images, specifically to tackle cell overlapping and imbalanced data. To separate the overlapping cells, our segmentation process first estimates ellipses of the RBC shapes. The method detects the concave points and then finds the ellipses using directed ellipse fitting on the set of detected convex curves. Our method achieves overlapping cell separation with the accuracy of 0.889. For RBC classification, we employ a deep learning technique, EfficientNet, and address the imbalanced training data problem with weight normalization, data up sampling, and focal loss.
Experimental results show that the weight balancing technique with augmentation has the best potential to deal with the imbalanced dataset problem by improving the F1-score on minority classes.  The best accuracy and F1-score are 0.921 and 0.8679, respectively, which outperform previous purposed models.
Code and dataset are available at https://github.com/Chula-PIC-Lab/Chula-RBC-12-Dataset.
\end{abstract}




\end{frontmatter}


\section{Introduction}
\label{sec:sample1}
A complete blood count (CBC) is a blood test that yields important information to evaluate the overall health and used to detect a variety of diseases. As a part of CBC, the red blood cell (RBC) morphology analysis must be performed. This analysis plays an essential role in diagnosing many diseases, caused by RBC disorders, such as anemia, thalassemia, sickle cell disease, etc. The analysis mainly focuses on the shape, size, color, inclusions, and arrangement of RBCs \cite{ford_red_2013}. The normal RBC shape is round, biconcave, with a pale central pallor and a diameter of 6–8 \textmu m. Generally, a hematologist manually analyzes the blood cells under a light microscope from blood smear slides. This manual inspection is a long process and also requires practices and experiences. Hence, here we employ the recent advancement of medical imaging and artificial intelligence-based (AI) technologies to provide an effective tool to help hematologists. The tool automatically and less subjectively analyze RBC images from a microscope, leading to reducing time and cost.

Deep convolutional neural networks for object detection and semantic segmentation have recently been employed for RBC detection \cite{zhang_rbc_2018, wong_analysis_2021, qiu_multi-label_2019, shakarami_fast_2021, DBLP:journals/corr/abs-1804-02767}. The benefits of these deep end-to-end methods are that, first, they enable training a possibly complex learning system represented by a single model, bypassing the intermediate layers usually found in the traditional pipeline approaches. Secondly, it is possible to design a model that performs well without deep knowledge about each sub-problem in the complex system.  However, the end-to-end approach is infeasible option in some cases, for example, a case where intermediate results are needed, and then the computational resources are not available.

The state-of-the-art deep learning methods of image classification [REFs] cannot be straightforwardly applied to identify the type of RBCs in the microscopic image. The biggest challenge is the overlapping cells as it is hard to find the edges of the cells, causing incorrect cell count results. 
The other challenge is the large amount of well-balanced data requirement. Although deep learning has shown remarkable results in computer vision, these approaches still need lots of data to achieve a good outcome. RBC datasets are difficult to collect because some RBC types can only be found in specific diseases, and these may only be found in specific geographic regions. Accordingly, each dataset usually has an highly imbalanced dataset problem. Moreover, different specialists might give different labels as their visual analysis is highly subjective.
Another challenge is system complexity. One of the main goal of this work is to develop the RBC segmentation and classification that can be implemented on the small devices, such as mobile phones and tablets, and demonstrate the classification results in real time. This will benefit in developing a mobile application for the hematologist training.   

In this paper, we propose a very efficient and accurate framework to overcome the challenges mentioned above. The method is based on traditional low-level computer vision to segment the overlapping cells.
 We proposed a new RBC segmentation method using ellipse fitting on the set of detected convex curves and classification using light-weight EfficientNet \cite{tan_efficientnet_2020}. The main contributions of the paper are: (i) a new method to separate the overlapping cells based on the concave points of the border of RBCs, and (ii) RBC classification on imbalanced datasets using data augmentation, weight normalization, upsampling, and focal loss \cite{lin_focal_2018} on multi-class classification.

\section{Total RBC counts and the automated methods}
Vision-based blood smear image analysis studies have been done on  specific cell type, such as white blood cell, RBC, and platelet \cite{hegde_peripheral_2018}. The automated counting and classification of RBCs used in CBC, or in specific disease diagnoses, such as malaria \cite{nugroho_feature_2015,liang_cnn-based_2016,gopakumar_convolutional_2018,poostchi_image_2018,delgado-ortet_deep_2020}, and anemia \cite{chandrasiri_automatic_2014,xu_deep_2017,abdulkarim_deep_2020,alzubaidi_deep_2020,yeruva_identification_2021} have been proposed. In this section, previous RBC studies are reviewed.
\subsection{RBC and its types}

\begin{figure}
    \centering
    \subfigure[Normal]
    {\includegraphics[width=1.1in]{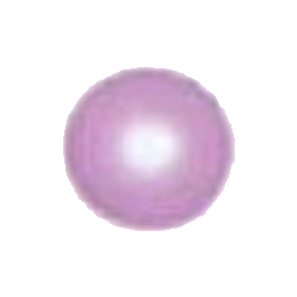}}
    \subfigure[Macrocyte]
    {\includegraphics[width=1.1in]{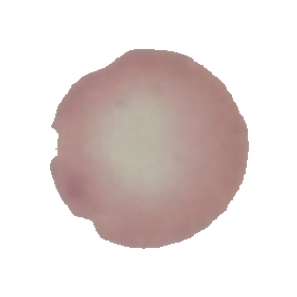}}
    \subfigure[Microcyte]
    {\includegraphics[width=1.1in]{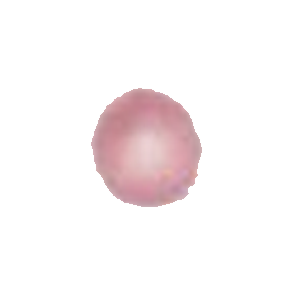}}
    \subfigure[Spherocyte]
    {\includegraphics[width=1.1in]{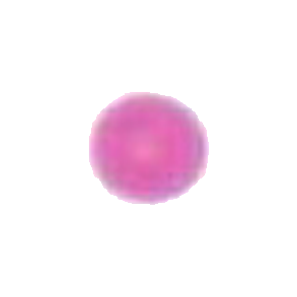}}
    \subfigure[Target cell]
    {\includegraphics[width=1.1in]{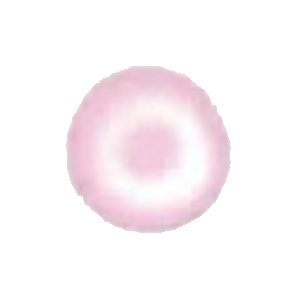}}
    \subfigure[Stomatocyte]
    {\includegraphics[width=1.1in]{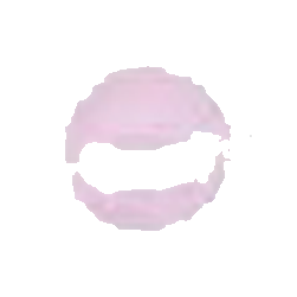}}
    \subfigure[Ovalocyte]
    {\includegraphics[width=1.1in]{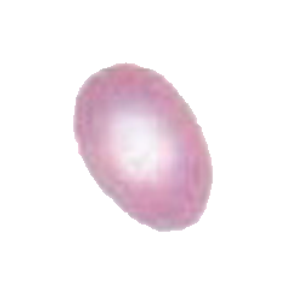}}
    \subfigure[Teardrop]
    {\includegraphics[width=1.1in]{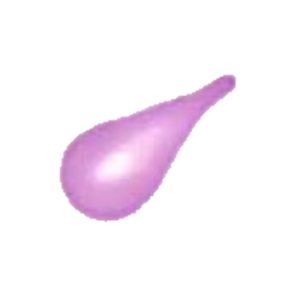}}
    \subfigure[Burr cell]
    {\includegraphics[width=1.1in]{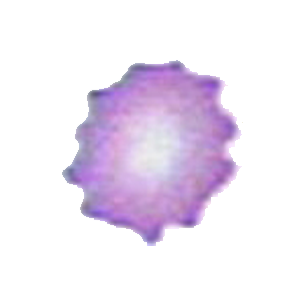}}
    \subfigure[Schistocyte]
    {\includegraphics[width=1.1in]{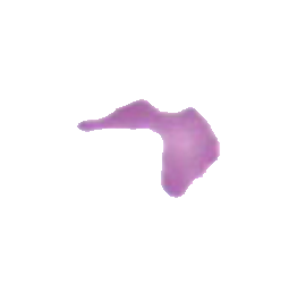}}
    \subfigure[Hypochromia]
    {\includegraphics[width=1.1in]{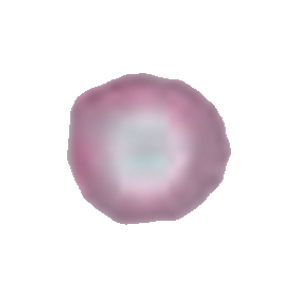}}
    
    \caption{Different types of RBCs}
    \label{fig:rbctypes}
\end{figure}

Figure \ref{fig:rbctypes} shows the eleven different types of RBCs that are focused in this study. As observed from a blood smear slide under a microscope, a normal RBC is a red circle with a white pale circle in the middle (the central pallor), and they are the main cells found in the blood. The other abnormal types are different from normal RBCs in shape, size, color, inclusion, and arrangement \cite{ford_red_2013}. For example, Macrocyte and Microcyte in Figure \ref{fig:rbctypes}(b, c) the RBCs are bigger and smaller than the normal, respectively, while Spherocyte, Target cell, Stomatocyte, and Hyprochromia in Figure \ref{fig:rbctypes}(d-f, k) are distinguished by observing their central pallor, and Ovalocyte, Teardrop, and Burr cell in Figure \ref{fig:rbctypes}(g-i) differ in the shape of the cells. Schistocyte in Figure \ref{fig:rbctypes}(j) shows a fragmented RBC. Note that inclusion RBCs, which are RBCs with dark dots inside, such as malaria parasite, are not included in this study. Although, a single RBC might exhibit more than one characteristic type, such as Oval + Macrocyte or Hypochromia + Macrocyte, our classification is formulated as a single-label multi-class classification, in which each RBC belongs to only one class.

\subsection{Segmentation of cells}
From our literature survey, we found that some of previous works performed only cell counting, while some works further analyzed and extracted shape information to used in subsequent classification steps. The automated process usually starts with converting the color image of blood smear images to a greyscale. Then, preprocessing is performed to improve the contrast and reduce the noise. Next, a segmentation step is applied followed by morphology operation to extract the RBC contours. To identify a number of RBCs, circle Hough transformation was used, since this works well for circular shaped RBCs \cite{mazalan_automated_2013,pasupa_convolutional_2020}. To extract the precise RBC shape, Otsu thresholding \cite{soltanzadeh_classification_2010,chandrasiri_automatic_2014,tomari_computer_2014,lee_cell_2014,romero-rondon_algorithm_2016,ahmad_geometrical_2016,acharya_identification_2017,gopakumar_convolutional_2018, aliyu_segmentation_2019} was used in many studies, to separate the RBC area from the background. Canny edge detection \cite{parab_red_2020,batitis_image_2020, alzubaidi_robust_2020} and Sobel edge detection \cite{rakshit_detection_2013} have been used. Other studies have used other methods, such as Dijkstra’s shortest path on the RBC border \cite{ritter_segmentation_2007}, ring shape dilation to identify the RBCs \cite{kareem_novel_2011}, and Watershed \cite{habibzadeh_counting_2011,sharma_detection_2016}, an active appearance model on sliding windows \cite{cai_red_2012}, K-means \cite{nugroho_feature_2015}, and histogram equalization \cite{tyas_classification_2017} have also been used. The most recent studies have used a fully convolutional neural network (FCN) \cite{delgado-ortet_deep_2020}, FCN-Alexnet \cite{sadafi_red_2018}, U-net \cite{de_haan_automated_2020}, and deformable U-net \cite{zhang_image_2017}. 

The result of segmentation might contain overlapping cells. To classify RBCs, most methods require separating them into single cells before computing their features or feeding into a deep learning model. \cite{sharma_detection_2016,pasupa_convolutional_2020} used the Circle Hough Transformation (CGT) to estimate each RBC’s contour.  Many works tried to identiy the number of RBCs in overlapping contour first, then separating the area of each cell.  Distance transformation was used to find a peak spot as a marker of each RBC; the Watershed  \cite{gopakumar_convolutional_2018} or a random walk method \cite{xu_deep_2017} was then applied. \cite{romero-rondon_algorithm_2016}proposed using K-means to cluster pixels of each cell, while \cite{gonzalez-hidalgo_red_2015} used ellipse adjustment with concave point finding approach.

From these previous works, extracting RBC regions from the background was not difficult because RBCs microscopic images usually have a high contrast.  However, separating cells that are connected to each other is still challenging because overlapping cells can be complicated and express in arbitrary shapes. The watershed method might yield good results but only when the cell centers are far apart.  The ellipse adjustment method relies on clear concave area between the cells.  The fully convolutional neural network (FCN) can detect the cell very accurate when trained by enough numbers of data with good quality labels. But it still cannot cope with the overlapping cells and must rely on the post processings.

\subsection{Classification of RBC types}
The early works in RBC classification were based on hand-crafted feature analysis and extraction, such as circularity, color average, diameter, area, etc. These features were used in classifiers based on rule-based methods \cite{soltanzadeh_classification_2010,chandrasiri_automatic_2014,acharya_identification_2017, batitis_image_2020,rahman_automatic_2021}, K-nearest neighbor \cite{sharma_detection_2016}, Support vector machine (SVM) \cite{vicent_algorithm_2022}, and artificial neural network (ANN) \cite{tomari_computer_2014,nugroho_feature_2015}, \cite{khashman_ibcis:_2008,tyas_classification_2017,lee_cell_2014,yeruva_identification_2021}. Later, convolutional neural networks (CNN) \cite{gopakumar_convolutional_2018,liang_cnn-based_2016,xu_deep_2017,durant_very_2017,alom_microscopic_2018,tiwari_detection_2018,delgado-ortet_deep_2020,alzubaidi_deep_2020,de_haan_automated_2020,abdulkarim_deep_2020} have been used to auto-generate features due to the high computing power available nowadays, and they can be extended to encompass more classes of RBCs. In recent classification studies, CNN classifiers have been used with more layers and new techniques. Durant et al. \cite{durant_very_2017} used DenseNet, which has more than 150 layers, to predict 10 RBC types. 

For end-to-end CNN approch, a U-net \cite{zhang_rbc_2018, wong_analysis_2021} have been used to perform both segmentation and classification in a single model. \cite{qiu_multi-label_2019} purposed multi-label detection using a Faster R-CNN with ResNet to detect six types of RBCs with touching and overlapping cells. \cite{song_red_2021} proposed attention residual feature pyramid network (ARFPN) to classify RBCs directly from the entire blood smear image. \cite{shakarami_fast_2021} used YOLOv3 \cite{DBLP:journals/corr/abs-1804-02767} to detect RBCs, WBCs, and platelets.  
From the review, most of the recent works were based on deep learning approaches which the accuracy depends on the size and quality of the dataset that was used. All previous study have used their own dataset in training and validating their models; this make it hard to compare the results. Moreover, each dataset contains small amounts of data, because the data collection is mostly done in a hospital, which is more difficult than general image dataset collection. For multi-class classification, some RBC types are rare and hard to find, and so the dataset may have an imbalance problem.

To classify such minority classes, imbalance handling techniques can help the classification model to be less biased towards the majority classes. Oversampling and undersampling are common techniques used in the data level analysis to deal with the imbalance problem. The simplest approach is data augmentation, where the existing data are used to generate more samples, through transformations such as cropping, flipping, translating, rotating and scaling \cite{AIreview:2021}. Another technique used in deep learning is modifying the training loss, where weights are applied differently to each class or the loss function is computed in region level, instead of pixel level \cite{8803305}.
Imbalance handling techniques on various data types at the data and algorithm levels have been reviewed \cite{johnson_survey_2019}, and shown that oversampling outperforms undersampling in most cases. Focal loss \cite{lin_focal_2018} is modified from the cross-entropy loss that helps the model to train on hard miss-classified samples and to focus less on well-classified samples. This technique has been employ in \cite{pasupa_convolutional_2020} to classify three RBC types from an imbalanced dataset. \cite{pasupa_semi-supervised_2020} used semi-supervised learning to train generative adversarial networks (GAN) to increase data samples.

\section{Proposed methods}
\begin{figure}[H]
    \includegraphics[width=\linewidth]{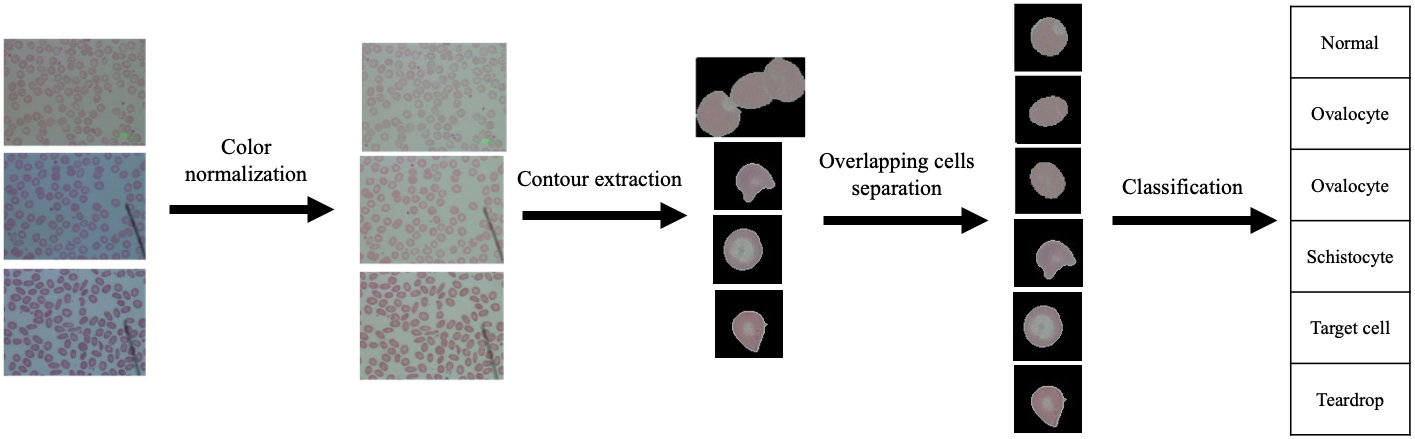}
    \caption{The pipeline of proposed method}
    \label{fig:overall}
\end{figure}

In this section, the proposed method is outlined, starting from normalizing the images and extracting an individual cell from blood smear images to identifying the type of each RBC, which is described by dividing it into the four main steps of: (i) RBC color normalization, (ii) overlapping cell separation, (iii) RBC contour extraction, and (iv) RBC classification. The overall process is summarized in Figure \ref{fig:overall}.

\subsection{Color normalization}

In the data collection process, collectors might have different environments, such as camera settings, microscope light levels, blood smear slide preparation, etc. The collectors also might collect multiple blood smear slides of a single RBC type at a time, making each type have its own color, as shown in Figure \ref{fig:overall}. Although, the human can disregard the difference in color seemlessly, the model can be biased by being trained with such data. Thus, the color normalization must be applied before using the images in training or inferencing by the model.

In our method, the backgrounds are extracted and the three overall average background values of the RGB channels were computed for all the blood smear images yielding $R_{avg} ,G_{avg}, B_{avg}$. To normalize the image k, the value difference of the three average background values of the target image $k$ ($r^k_{avg} ,g^k_{avg}, b^k_{avg}$) and the overall averages values were added to all the pixels of the target image. The normalization equation of pixel $(i,j)$ for image $k$ is shown in Eq. \ref{normalize_equation}. The result shown in Section 5.2 shows that the normalization improves the classification while showing no negative effect to the segmentation.

\begin{equation}
\label{normalize_equation}
\begin{aligned}
    r'^k_{i,j} = r^k_{i,j}+(R_{avg}-r^k_{avg})\\
    g'^k_{i,j} = g^k_{i,j}+(G_{avg}-g^k_{avg})\\
    b'^k_{i,j} = b^k_{i,j}+(B_{avg}-b^k_{avg})
\end{aligned}
\end{equation}

\subsection{RBC contour extraction}
\begin{figure}
    \centering
    \subfigure[]
    {
        \includegraphics[width=1.3in]{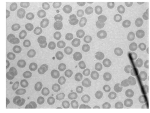}
    }
    \\
    \subfigure[]
    {
        \includegraphics[width=1.3in]{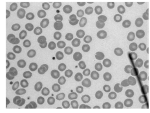}
    }
    \subfigure[]
    {
        \includegraphics[width=1.3in]{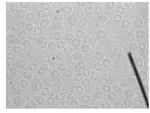}
    }
    \caption{The RGB channels: (a) red channel image, (b) green channel image, and (c) blue channel image}
    \label{fig:rgbimage}
\end{figure}

\begin{figure}
    \centering
    \subfigure[]
    {
        \includegraphics[width=1.3in]{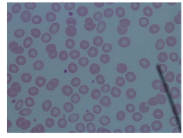}
    }
    \subfigure[]
    {
        \includegraphics[width=1.3in]{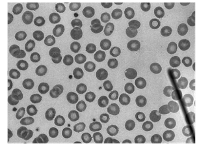}
    }
    \\
    \subfigure[]
    {
        \includegraphics[width=1.3in]{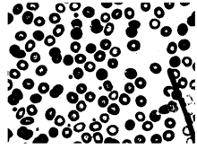}
    }
    \subfigure[]
    {
        \includegraphics[width=1.3in]{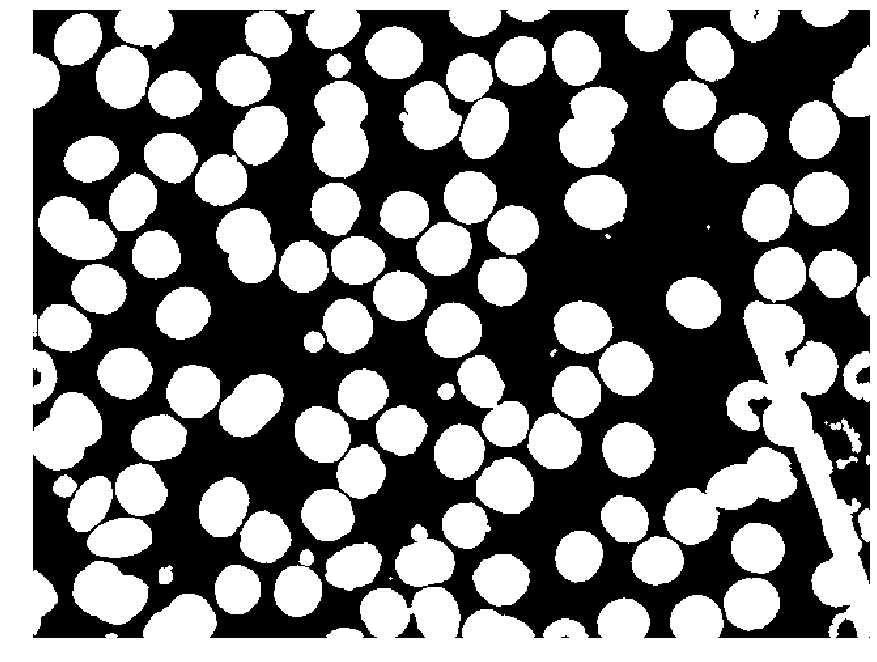}
    }
    \caption{The (a) original image, (b) CLAHE image, (c) threshold image, and (d) contour image}
    \label{fig:countour}
\end{figure}

To extract the RBC contours from blood smear images, from the input color image, we select only the green channel because it has more contrast than other channels (Figure \ref{fig:rgbimage}). Then, CLAHE (Contrast Limited Adaptive Histogram Equalization) is used to enhance the RBC regions making them stand out of the background. Finally, the Otsu thresholding and contour findings are used to extract cell regions. The overall process is shown in Figure \ref{fig:countour}.

\subsection{Overlapping cell separation}
In manual RBC analysis, hematologists typically avoid selecting an area in a blood smear slide that has overlapping cells to evaluate the result. This is because it is simple to count and identify the type of RBC when their border is not hidden behind other cells. To separate the overlapping RBCs, the most reliable methods are based on distance transforming and ellipse fitting. The distance transform approach is used to find the peak spot furthest from the border pixels. The peak spots are then used to identify a unique cell by several technique, such as the random walk method or watershed transform. However, the distance transform works effectively only on the circular shaped object and partially overlapped cells. When the cells mostly overlap to each other, the distance transform might not be able to distinguish the peak for each cell. The ellipse fitting method uses the edge of the RBC to approximate as an ellipse which identifies the area of RBC.

The method presented herein is based on ellipse fitting and the overall process was divided into four steps, as detailed below and shown in Figure \ref{fig:overlappingStp}.

\subsubsection{Concave point finding} 
Many features have been used to identify the overlapping cells such as the size of the RBC, the circularity of RBCs, and the contour of the overlapped RBC area. In this study, the contour is used due to its local characteristic comparing to the other two which use the global characteristic in detecting each cell from the overlapped area. Using contour analysis and finding concave points along the contour allow us to estimate the precise shape of the RBCs. The edge of the contour between two concave points can be identified as the edge of a single cell. Moreover, the edge between two concave points also can be used to identify the number of RBCs.

\begin{figure}
    \includegraphics[width=\linewidth]{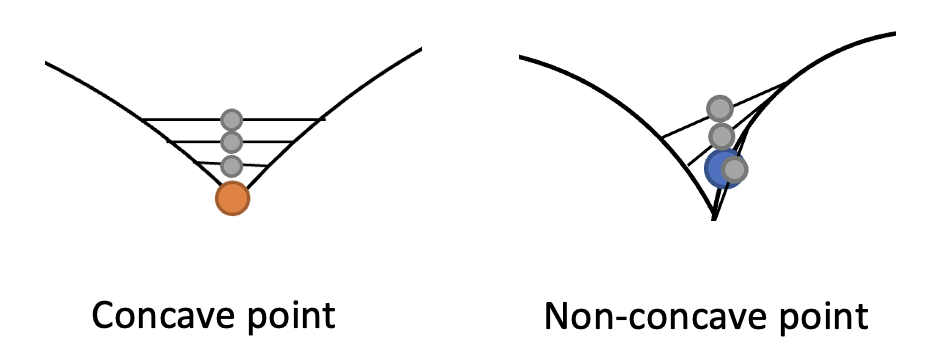}
    \caption{The illustration of the concave point finding. The orange point is a concave point because all $k$ middle points (the gray points) are outside the contour. The blue point is a non-concave point because at least 1 middle point is inside the contour.}
    \label{fig:concavepoint}
\end{figure}

To find the concave points, our algorithm runs as follows: for each boundary pixel ($x_i$,$y_i$) on the contour of overlapping region, $k$ middle points, which locate at the center of the $k$ pairs of left and right neighboring pixels of ($x_i$,$y_i$), were calculated. If all $k$ points are outside the contour, the point is considered as a concave point. Figure \ref{fig:concavepoint} illustrates how we find the concave points. Instead of checking just one pair, checking $k$ pairs makes the method more robust because the edge of RBCs may not smooth and can introduce the false positive concave points. Increasing the number of pairs of contour points can decrease the false positive concave points without using the edge smoothing technique which can increase the complexity of the process. Occasionally, we might find the contiguous concave points. In such a case, we will select only the middle one. The concave points function, $f(x)$, is calculated using Eqs. (2) and (3):
\begin{equation}
    f(x_i,y_i) = \prod_{j=1}^k{g(\frac{x_{i-j}+x_{i+j}}{2},\frac{y_{i-j}+y_{i+j}}{2})}
\end{equation}
\begin{equation}
    g(x) = \begin{cases}1 , (x,y) &\text{is outside a contour,} \\
    0 , (x,y) & \text{is inside a contour}\end{cases}
\end{equation}

\subsubsection{Ellipse fitting}
If the contour has more than one concave point, convex curves between each pair of consecutive concave points are used to approximate an ellipse shape by direct ellipse fitting \cite{fitzgibbon_direct_1999}, based on the least-square method. The direct ellipse fitting is recommended instead of the original \cite{fitzgibbon_buyers_1995}, which gives an approximate ellipse that does not relate to the curve in some conditions. The direct ellipse fitting is constrained by ensuring the discriminant $4ac-b^2 =0$ for the ellipse equation, as shown in Eq. (\ref{eq:elipse}).
\begin{equation} \label{eq:elipse}
    ax^2+bxy+cy^2+dx+ey+f=0
\end{equation}
\subsubsection{Ellipse verification}
After finding all the ellipses in each overlapping contour, the ellipses are sorted by area in descending order. Then, each ellipse is verified whether it is a RBC by checking with three criteria: (i) 80\% of the ellipse area is in the overlapping contour, (ii) at least 20\% of the ellipse area must not be in any previous ellipses., and (iii) the size of the ellipse must be greater than the size filtering threshold we used in filtering out the noise and small objects. Large ellipses were considered first because RBCs should cover the major part of the overlapping contour and these ellipses are mostly on the wider curve. Small ellipses are usually on short curves and may not cover the correct shape of the cells so they can be eliminated by the conditions (ii). Figure \ref{fig:overlappingStp} (c) shows the correct (green ellipses) and incorrect (blue ellipses) ellipse estimations.
\subsubsection{Two curve ellipse fitting}
\begin{figure}
    \centering
    \subfigure[]
    {\includegraphics[width=1.in]{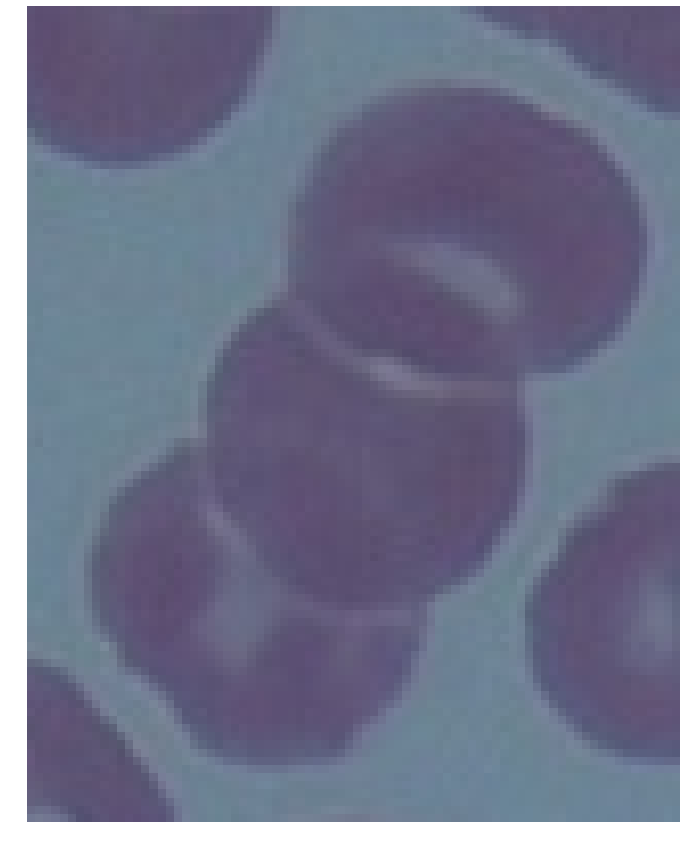}}
    \subfigure[]
    {\includegraphics[width=1.in]{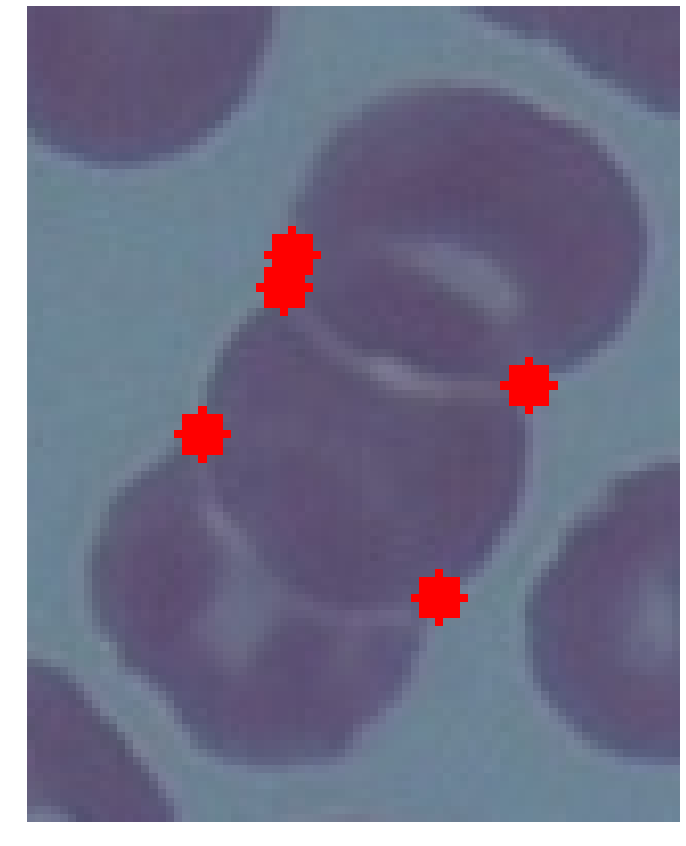}}
    \subfigure[]
    {\includegraphics[width=1.in]{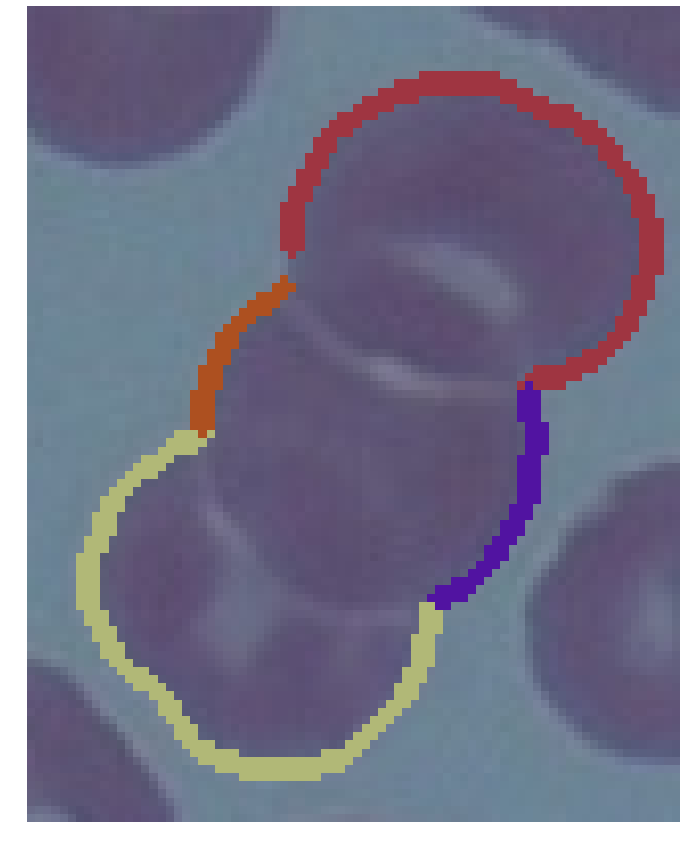}}
    \subfigure[]
    {\includegraphics[width=1.in]{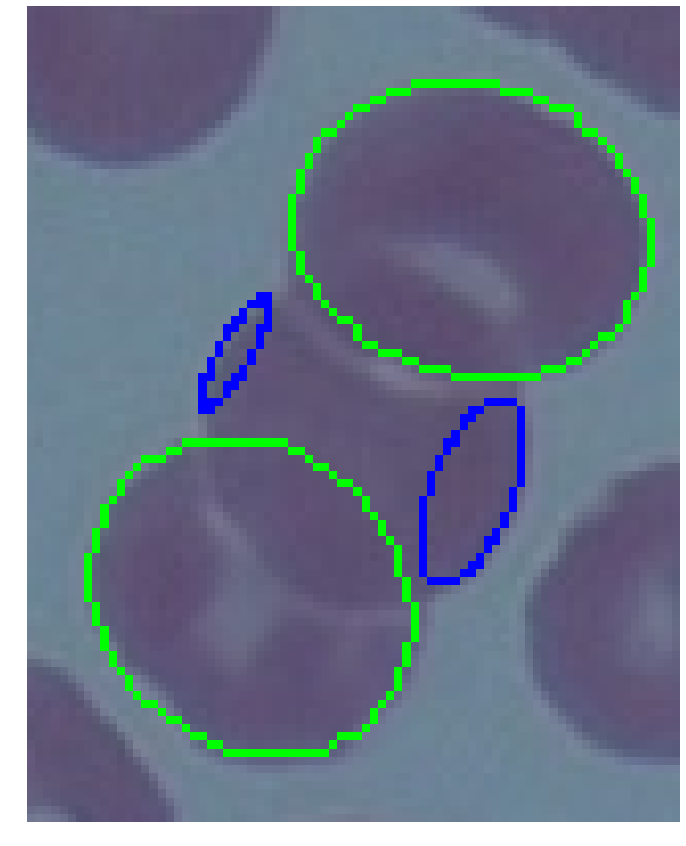}}
    \subfigure[]
    {\includegraphics[width=1.in]{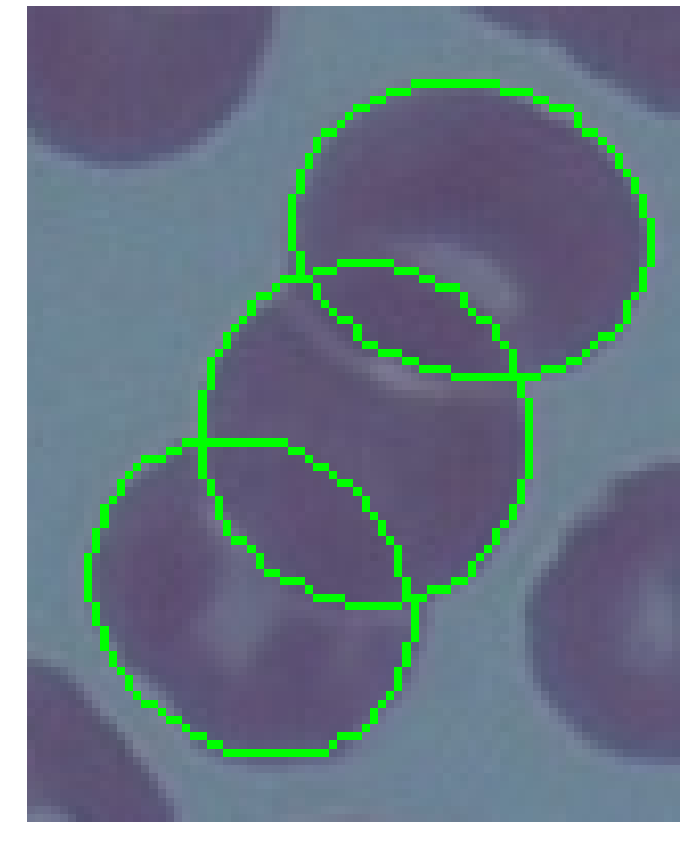}}
    
    \caption{Overlapping cell separation steps: (a) input image (b) five concave points are detected, but only four points are selected using proximity analysis. (c) two consecutive concave points form a convex curve yielding 4 convex curves in this example. (d) ellipse fitting is performed on each convex curve yielding 4 ellipses representing RBCs. The green ones pass the ellipse verification criteria, while the blue ones fail. Thus they are merged and being fit again yielding a new ellipse that eventually meet all three criteria (e).}
    \label{fig:overlappingStp}
\end{figure}

In some complicated overlapping, i.e., more than two cells overlap each other, some curves especially the small ones might not able to restore the correct ellipse shape for RBC as shown in Figure \ref{fig:overlappingStp}(d). We will combine two remaining curves as inputs for ellipse fitting. This two-curve ellipse fitting is performed on all combination of the remaining curves and then verified using the three criteria mentioned previously to yield the final result.

\subsection{RBC classification}
\subsubsection{RBC dataset}
The dataset of this work was collected at the Oxidation in Red Cell Disorders Research Unit, Chulalongkorn University. 706 blood smear images were acquired using a DS-Fi2-L3 Nikon microscope at 1000x magnification. The images were then segmented to individual cells and labeled by specialists in hematology. The dataset contain 12 classes of RBCs including 11 RBC types as shown in Figure 1 and an uncategorized type. The numbers of cells in each class are shown in Table \ref{tab:total}. The dataset is highly imbalanced because some classes such as Microcyte and Teardrop are rare.

\begin{figure}
    \includegraphics[width=\linewidth]{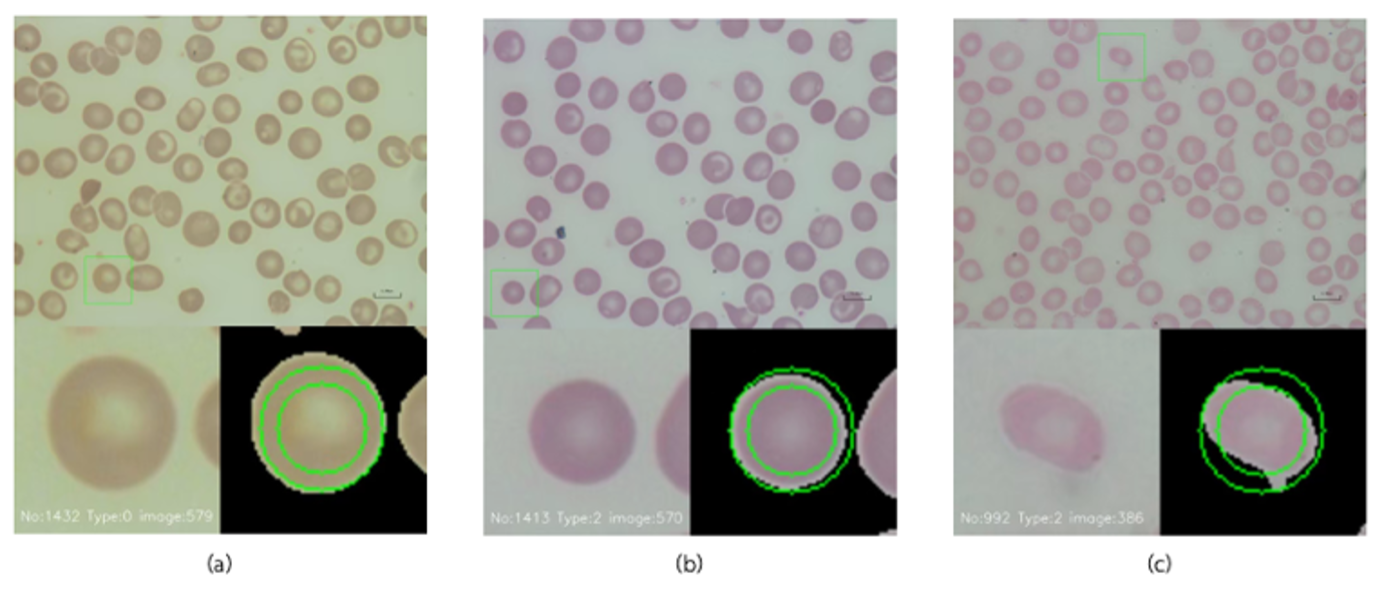}
    \caption{Examples of inconsistent labeling. (a) Macrocyte was labeled as Normal cell, (b) Normal cell was relabeled as Microcyte, (c) Ovalocyte was labeled as Microcyte. To cope with this, a measurement tool was developed and majority vote was also applied.}
    \label{fig:rebalel}
\end{figure}

In cell labeling, labelers can disagree in classifying cells especially the ones with scale difference and ones with ambiguous shape.  To cope with this problems, we developed a tool to help labelers to measure the size of the cells (Figure \ref{fig:rebalel}).  Also, majority vote is applied when the labelers disagreed.

\begin{table}
\caption{Summary of our dataset}
\centering
\begin{tabular}{|l|r|}
\hline
RBC Class&
Number of samples\\
\hline
Normal&6,286\\
Macrocyte&687\\
Microcyte&459\\
Spherocyte&3,445\\
Target cell&2,703\\
Stomatocyte&1,991\\
Ovalocyte&2,137\\
Teardrop&305\\
Burr cell&783\\
Schistocyte&861\\
Hypochromia&1,036\\
Uncategorised&182\\

\hline
Total&20,875\\
\hline
\end{tabular}
\label{tab:total}
\end{table}

\subsubsection{EfficientNet}
For classification, the pretrained EfficientNet model \cite{tan_efficientnet_2020} was used as it showed a remarkable level of accuracy and better performance than the older models. It was designed by carefully balancing the network depth, width, and resolution. The model has eight different sizes: EfficientNet-B0 to EfficientNet-B7. In the results section, the EfficientNet-B0 to EfficientNet-B4 were observed with a five-fold cross validation using 80\% and 20\% for training and testing, respectively. Moreover, we used data augmentation and weight initialization to observe which technique could overcome the class imbalance problem. Only random flips and rotates were used for data augmentation because the RBC classes are sensitive to size and color, such as Normal Macrocytes and Microcytes are different in size.

To evaluate the performance, accuracy is commonly used for image classification. For the imbalanced dataset, the accuracy is insufficient, as it can be dominated by the majority classes. However, many metrics have been used to describe an imbalanced dataset \cite{johnson_survey_2019}, and the F1-score was used in this study. This is a well-known metric that balances precision and recall by harmonic means that is sensitive to the minority classes.

\subsubsection{Imbalance handling techniques}
An imbalanced dataset is a common problem in biomedical datasets. Our RBC dataset is highly imbalanced with a 34.538 imbalance ratio (a proportion samples of majority class to the minority class) for the 12 RBC classes of 20,875 RBCs. In the training step, the model can be overcome by high sample classes with less focus on the low sample classes. Thus, weight balancing, up sampling, and focal loss were investigated in this study.

For weight balancing, normally, every RBC class has the same weight, $1.0$. However, the weight balancing helps a model balance learning gradients in the backpropagation step between high sample classes and low sample class, by giving a high weight to low sample classes and a low weight to high sample classes. In this study, each class was weighted as $\frac{1}{f}$, $\frac{1}{\sqrt{f}}$, and $\frac{1}{\sqrt[3]{f}}$  as Weight, Weight2, and Weight3, respectively, where $f$ is the amount of samples in that class.

The up sampling makes every RBC class have the same amount of samples by replicating its own data. This helps the trained model to not be overcome by high sample classes. In this case, every class  replicates itself to match the normal class.

Focal loss is used to help the model focus on the high loss and reduce the loss near 0. It was used in the object detection task, which is highly imbalanced between objects and nonobject classes. We investigated if multi-class classification could help in an imbalanced multi-classification. As shown in Eq. \ref{eq:focal}, focal loss has an added term from the cross entropy loss, which reduces the loss when the predicted probability result is close to the known truth by the $\gamma$ hyperparameter.
\begin{equation}
    FL(p_t) = -(1-p_t)^\gamma log(p_t)
    \label{eq:focal}
\end{equation}

\section{Implementations}

This section gives all the parameter values we used in our implementation in this study. The blood smear image size in our dataset is $640 \times 480$ pixels. For the overlapping cells separation, the $k$ value was 8 which can reduce the false-positive concave point for our image resolution. The RBC contours were cropped and put on a blank $72 \times 72$ image, which is big enough to contain the largest cell.

For classification, the contour images were scaled to $224 \times 224$, which is the input size of the EfficientNet model. The models were trained with pretrained parameters from ImageNet. The learning rate was 0.001 with a 0.1 learning decay rate in every 15 epochs, and a total of 100 epochs.

\section{Results}
In this section, the overlapping cells separation and RBC classification results are provided and analyzed.

\subsection{Overlapping cell separation}
\begin{figure*}
    \centering
    {\includegraphics[width=2in]{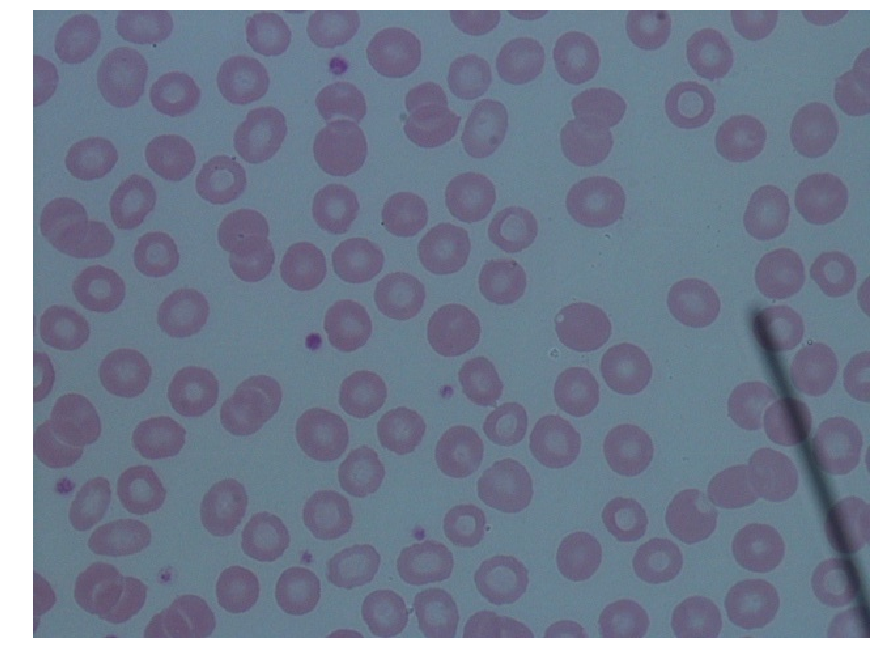}}
    {\includegraphics[width=2in]{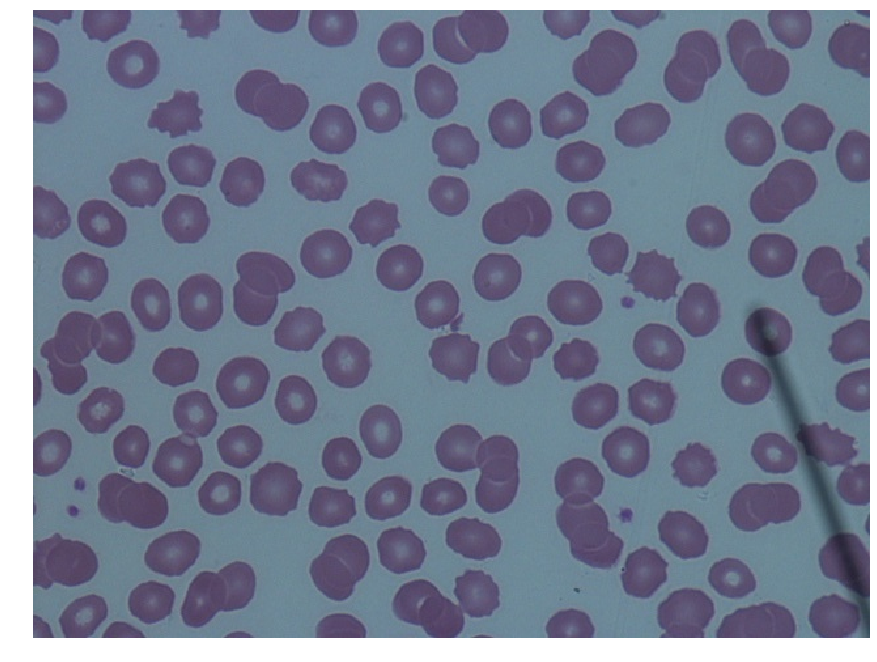}}
    {\includegraphics[width=2in]{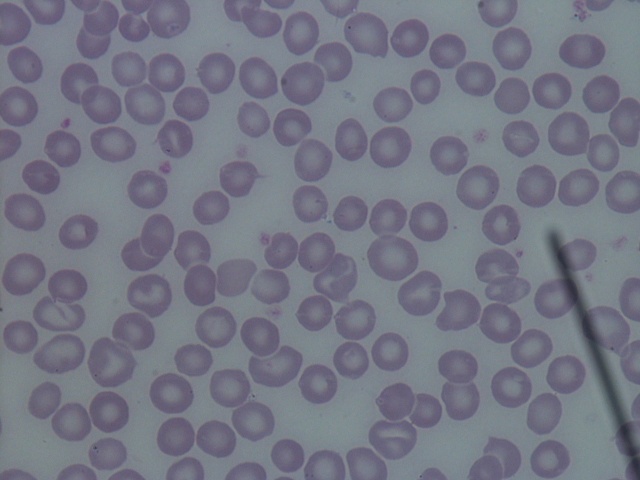}}\\
    {\includegraphics[width=2in]{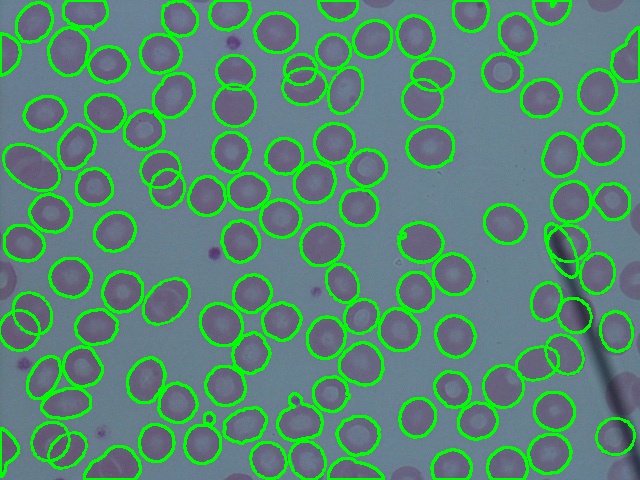}}
    {\includegraphics[width=2in]{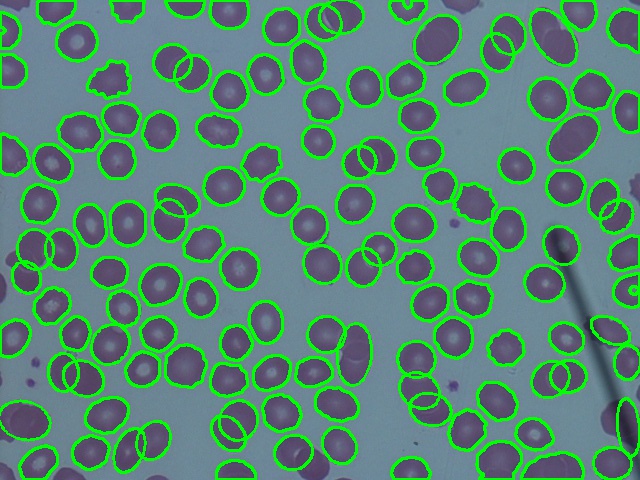}}
    {\includegraphics[width=2in]{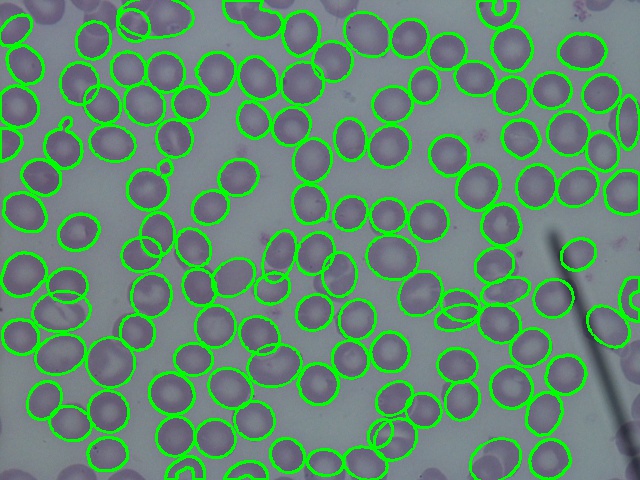}}

    \caption{Segmentation results}
    \label{fig:rbcsegmentresult}
\end{figure*}

To evaluate the separation of the overlapping cells, we manually counted the overlapping contours that exclude cells on the border of the images and other artifacts, such as platelets, white blood cells, or microscope tools in the images. A total of 277 contours were found in 20 blood smear images, with an overall accuracy of 0.889. Most of the contours were two RBCs touching or overlapping with each other. The error in this method was mostly found on contours that had only one concave point. The results are summarized in Table \ref{tab:concaveresult}, while blood smear images after segmentation are shown in Figure \ref{fig:rbcsegmentresult}.

\begin{table}
\caption{Overlapping cells separation results}
\fontsize{9}{9}
\centering
\begin{tabular}{|l|c|c|c|c|}
\hline
Contour&
Correct&
Incorrect&
Incorrect&
Total\\
&&(Concave)&(Fitting)&\\
\hline
2 RBCs&	185&	18&	4&	207\\
3 RBCs&	38&	2&	1&	41\\
$>$3 RBCs&	23&	2&	4&	29\\
\hline
Total&	246&	22&	9&	277\\
\hline
\end{tabular}
\label{tab:concaveresult}
\end{table}

\subsection{RBC classification}
In the first step, we investigated the different model sizes, EfficientNet-B0 to B4, with and without augmentation. The results (Table \ref{tab:classificationmodelsize}) show that EfficientNet-B1 with augmentation had the highest accuracy and F1-score. Thus, increasing the model size did not significantly improve the performance, and the limiting factor was the sample size of the dataset. Increasing the model size can then lead to an overfitting problem. Therefore, imbalance handling techniques were investigated, including weight balancing, up sampling, and focal loss, using EfficientNet-B1 as the baseline.

\begin{table}
\caption{RBC classification results}
\centering
\begin{tabular}{|l|c|c|}
\hline
Model&
Accuracy&
F1-score\\
\hline
EfficientNet-B0&	0.8821&	0.8378\\
EfficientNet-B1&	0.8823& 0.8426\\
EfficientNet-B2&	0.8842& 0.8399\\
EfficientNet-B3&	0.8819& 0.8423\\
EfficientNet-B4&	0.8830& 0.8405\\
\hline
EfficientNet-B0-aug&	0.8996&	0.8639\\
\textbf{EfficientNet-B1-aug}&\textbf{0.9021}	& \textbf{0.8679}\\
EfficientNet-B2-aug&	0.8988& 0.8636\\
EfficientNet-B3-aug&	0.9001& 0.8642\\
EfficientNet-B4-aug&	0.8990& 0.8668\\
\hline
\end{tabular}
\label{tab:classificationmodelsize}
\end{table}

The overall training accuracy and F1-score of EfficientNet-B1 with imbalance handling techniques are summarized in Table \ref{tab:classificationb1}. However, the baseline model with augmentation still had the highest accuracy and F1-score, followed by AugWeight3 (augmentation and $\frac{1}{\sqrt[3]{f}}$ weight). Up (Up sampling) showed a slightly lower result from the baseline while AugUp (Augmentation with up sampling) showed slightly better results. Augmentation with focal loss (AugFocal0.1-AugFocal3.0) resulted in a decreasing accuracy and F1-score with increasing $\gamma$ hyperparameter values.

\begin{table}
\caption{Accuracy and F1-Score of our proposed EfficientNet-B1 with various data imbalance handling techniques}
\centering
\begin{tabular}{|l|c|c|}
\hline
Model&
Accuracy&
F1-score\\
\hline
Baseline&	0.8823&	0.8426\\
\hline
Aug&\textbf{0.9021}	& \textbf{0.8679}\\
\hline

Weight&	0.8752&	0.8374\\
Weight2&	0.8808&	0.8435\\
Weight3&	0.8820&	0.8410\\
AugWeight&	0.8698&	0.8344\\
AugWeight2&	0.8954&	0.8630\\
AugWeight3&	0.8981&	0.8672\\
\hline

Up&0.8772&	0.8403\\
AugUp&0.8877&	0.8591\\
\hline
AugFocal0.1&0.8965&	0.8586\\
AugFocal0.2&0.8977&	0.8584\\
AugFocal0.3&0.8963& 0.8594\\
AugFocal0.4&0.8966& 0.8589\\
AugFocal0.5&0.8947&	0.8523\\
AugFocal1.0&0.8932&	0.8510\\
AugFocal1.5&0.8926&	0.8543\\
AugFocal2.0&0.8900&	0.8480\\
AugFocal2.5&0.8884&	0.8486\\
AugFocal3.0&0.8877&	0.8488\\
\hline

\end{tabular}
\label{tab:classificationb1}
\end{table}

\begin{figure}
    \includegraphics[width=\linewidth]{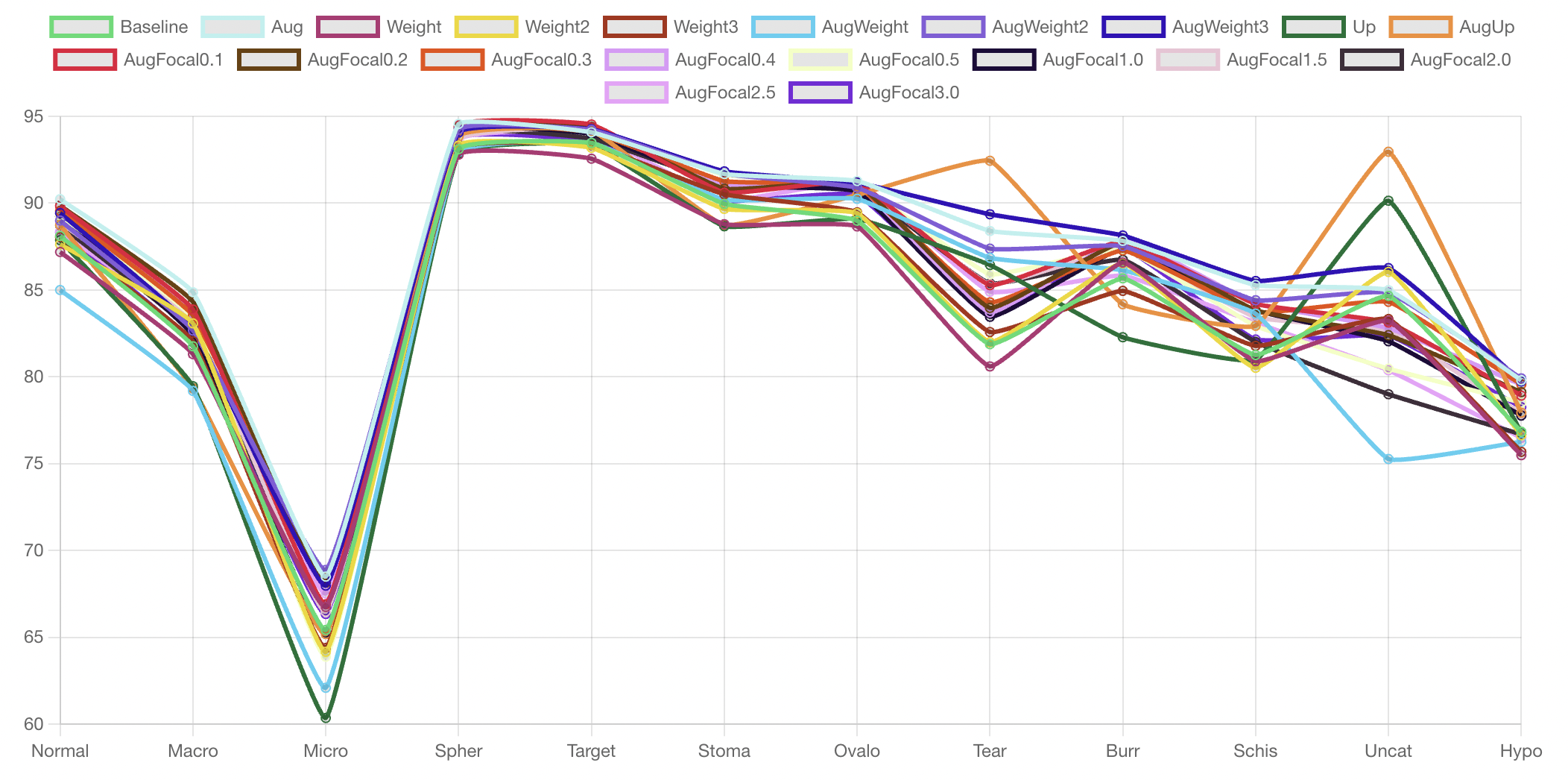}
    \caption{F1-score of classifying RBCs classes using EfficientNet-B1 with data imbalance handling techniques}
    \label{fig:linechart}
\end{figure}

Table \ref{tab:classificationb1vlass} and Fig \ref{fig:linechart} summarize the F1-score in each class with data imbalance handing techniques. The results revealed that Aug improved all class F1-scores from the baseline. Weight - Weight3 and AugWeight - AugWeight3 showed similar results compared to the baseline and Aug, respectively. AugWeight3 had six classes of RBCs with a better F1-score than Aug, which implied that an extremely different weight might not give an accurate result. AugUp had three RBC classes that were better than Aug with a huge improvement in the recognition of the Teardrop and Uncategorized types of RBCs. The focal loss was decreased when increasing the $\gamma$ hyperparameter in all RBC classes.

In model training, the weight balancing, augmentation and focal loss do not increase the number of sample data but the up sampling does. It increases the number of data samples in each class to match the number of the Normal class samples. This makes the dataset 6-fold increase, and thus takes much more time in training. For training with the focal loss, although the training data not increase, but the grdient was reduce because of the $\gamma$ hyperparameter values thus making the time for training slignly longer than the others.

\begin{table*}
\caption{F1-score of classifying RBCs classes using EfficientNet-B1 with data imbalance handling techniques}
\centering
\begin{adjustbox}{width=\textwidth}

\begin{tabular}{|l|c|c|c|c|c|c|c|c|c|c|c|c|}
\hline
Techniques& Normal& Macro& Micro& Spher& Target& Stoma& Ovalo& Tear& Burr& Schis& Uncat&Hypo\\ \hline
Baseline& 88.11& 81.76& 65.43& 93.11& 93.47& 89.96& 88.99& 81.88& 85.67& 81.26& 84.66& 76.84\\
Aug& \textbf{90.23}& \textbf{84.87}& 68.50& \textbf{94.56}& 94.08& 91.68& \textbf{91.28}& 88.40& 87.83& 85.28& 84.99& 79.80\\
Weight& 87.19& 81.31& 66.75& 92.79& 92.56& 88.79& 88.66& 80.60& 86.58& 80.90& 83.23& 75.49\\
Weight2& 87.63& 83.08& 64.17& 93.31& 93.19& 89.67& 89.42& 82.00& 86.54& 80.52& 86.04& 76.64\\
Weight3& 88.04& 82.05& 64.39& 93.12& 93.33& 90.51& 89.47& 82.58& 84.95& 81.78& 83.33& 75.68\\
AugWeight& 85.00& 79.19& 62.10& 92.95& 93.48& 90.20& 90.25& 86.85& 86.14& 83.64& 75.27& 76.26\\
AugWeight2& 89.00& 82.67& \textbf{68.88}& 94.23& 94.22& 91.69& 90.89& 87.38& 87.53& 84.41& 84.85& \textbf{79.91}\\
AugWeight3& 89.43& 82.82& 67.98& 94.07& 94.26& \textbf{91.83}& 91.18& 89.36& \textbf{88.14}& \textbf{85.52}& 86.26& 79.74\\
Up&87.87&	79.46&	60.36&	92.86&	93.36&	88.67&	89.10&	86.44&	82.28&	80.98&	90.14&	76.81\\
AugUp&88.74&	79.34&	65.19&	93.85&	94.17&	88.73&	90.47&	\textbf{92.44}&	84.18&	82.94&	\textbf{92.96}&	77.94\\
AugFocal0.1&89.83&83.88&66.91&94.47&\textbf{94.53}&90.61&91.20&85.28&87.80&84.20&83.13&78.92\\
AugFocal0.2&89.86&84.32&68.15&94.35&94.40&90.83&91.11&83.97&87.74&83.83&82.40&79.12\\
AugFocal0.3&89.56&83.53&68.43&94.20&94.09&91.27&90.97&84.29&87.28&83.72&84.32&79.57\\
AugFocal0.4&89.45&83.65&68.26&94.25&94.37&91.09&91.00&83.71&87.93&84.22&82.81&79.55\\
AugFocal0.5&89.28& 83.37& 63.89& 93.94& 94.30& 91.62& 91.12& 85.91& 87.36& 82.93& 80.47& 78.52\\
AugFocal1.0&89.25& 82.37& 65.29& 94.09& 93.99& 91.01& 90.67& 83.46& 87.38& 83.83& 82.05& 77.77\\
AugFocal1.5&88.89& 83.10& 66.48& 93.68& 94.24& 91.01& 91.14& 85.32& 87.34& 83.43& 82.64& 77.88\\
AugFocal2.0&88.73& 82.50& 66.89& 93.95& 93.72& 91.12& 90.86& 85.38& 86.74& 82.01& 79.00& 76.69\\
AugFocal2.5&88.37& 82.72& 67.53& 93.66& 94.01& 90.15& 91.18& 84.90& 85.83& 83.12& 80.38& 76.49\\
AugFocal3.0&88.34& 81.99& 66.35& 93.71& 93.55& 90.14& 90.48& 83.76& 87.41& 82.15& 82.38& 78.25\\

\hline
\end{tabular}
\end{adjustbox}
\label{tab:classificationb1vlass}
\end{table*}

In summary, the difference between this dataset and general datasets in the RBC classification problem is the dataset is imbalanced and RBC classes have many similar characteristics. Almost all classes are circular in shape, with only a few characteristics that are different, such as their size, shape, and color. The best result was obtained with the EfficientNet-B1 with augmentation.

Further analysis on an imbalanced dataset, weight balancing, and focal loss was examined for their effect on the loss function. Weight balancing helped to improve the low sample classes with less focus on the high sample classes. Otherwise, focal loss showed a decreased performance for this dataset because it focused on a high value loss, but since the different RBC classes were almost similar in shape the loss was almost entirely in the middle, which is ignored. Up sampling was performed at the data level, similar to augmentation. This technique seemed to work best for unique shape classes, which were the teardrop cell and uncategorized classes.

The analysis with a normalization step (Table \ref{tab:classificationb1alter}) revealed a huge improvement compared to augmentation with an unnormalized image. We also trained the model with the different background colors of black, white, grey, and the average background color, where the black background gave the best result, slightly higher than the others.

    To compare the performance with different models, we compared the accuracy and F1-score with ResNet and DenseNet which were used in previous RBC classification works \cite{pasupa_convolutional_2020,durant_very_2017} and MobileNetV3 \cite{howard_searching_2019} which was designed for running in mobile phone CPU. The MobileNetV3-Small, MobileNetV3-Large, ResNet50, ResNet121, DenseNet-121, and DenseNet-169 were trained on our dataset with augmentation technique. The results are shown in Table \ref{tab:classificationcompare}. Our model outperforms the previous model on both accuracy and F1-score. However, the result of EfficientNet-b1 is only slightly better than the previous models. According to the results, all models seem to overfit to our dataset.

Recently, \cite{wong_analysis_2021} reported the results of using SVM and TabNet to classify the RBCs into 11 classes on the same dataset as used in this paper.  They employed the SMOTE technique with cost-sensitive learning to handle the imbalanced dataset.  The evaluation was done using F2-Score and the results show that the SVM outperforms the TabNet with 78.2\% and 73.0\% respectively.  To compare with their work, we employed our methods, EfficientNet-B1 with augmentation, to classify 11 and 12 classes of RBCs on the same dataset.  Our approach yields 88.62\% and 87.91\% F2-Score respectively.


\begin{table}
\caption{Accuracy and F1-score of EfficientNet-B1 with different normalization techniques}
\centering
\begin{tabular}{|l|c|c|}
\hline
Model&
Accuracy&
F1-score\\
\hline

AugUnnormalize&0.6325&	0.4241\\
\hline
AugBlackbg&\textbf{0.9021}	& \textbf{0.8679}\\
AugWhitebg&0.8977&	0.8634\\
AugGraybg&0.8969&	0.8603\\
AugAVGbg&0.8979&	0.8626\\
\hline
\end{tabular}
\label{tab:classificationb1alter}
\end{table}

\begin{table}
\caption{Comparison on accuracy, F1-score, and number of parameters of the previous models that were trained on our dataset}
\centering
\begin{tabular}{|l|c|c|c|}
\hline
Model&
Accuracy&
F1-score& 
Number of parameters\\
\hline

EfficientNet-b1&\textbf{0.9021}	& \textbf{0.8679}& 7.8M\\
\hline
MobileNetV3-Small&0.8923&0.85830&2.9M\\
MobileNetV3-Large&0.8948&0.8586&5.4M\\
DenseNet-169&0.8990	& 0.8675&14M\\
DenseNet-121&0.8981&	0.8678&7.2M\\
ResNet152&0.8983&	0.8661&60M\\
ResNet50&0.8974&	0.8651&26M\\

\hline
\end{tabular}
\label{tab:classificationcompare}
\end{table}

\subsection{Time performance}
The motivation behind this work is that we want to develop an automatic RBC detection and classification application on mobile devices. Thus, the efficiency is one of our concerns when designing the algorithm. In this section, the time that was used for each step including segmentation, overlapping cell separation, and classification were measured on a personal computer with Intel Core(TM) i7-4770 3.40GHz CPU, 16 gigabyte RAM, and NVIDIA GTX 1080 Ti GPU.

The average time of each step was measured on 20 blood smear images, as shown in Table \ref{tab:timeeval}. The segmentation used only 0.032 seconds and the overlapping cell separation used 0.737 seconds; this make it feasible to run on edge devices such as a smartphone in real-time. However, the algorithm can be optimized by running in parallel. For the classification, at test time, RBCs were fed into the model one at a time. The GPU can run almost 3 times faster than the CPU. It is might not suitable to run on edge devices. To improve the running time, a smaller model and trade-off between accuracy and running time may be considered. Table \ref{tab:classificationcompare} shows the number of parameters of each model. Our model has 7.8 million parameters which only slightly higher than DenseNet-121 but it was yields better in accuracy and F1-score.

\begin{table}
\caption{The average time measurements of segmentation, overlapping cell separation, and classification processes}
\centering
\begin{tabular}{|l|c|}
\hline
Process&
Time (Seconds)\\
\hline
Segmentation&0.0320\\
\hline
Overlapping cell separation&0.7370\\
\hline
Classification (CPU)&7.5183\\
Classification (GPU)&2.7796\\
\hline
Total (CPU)&8.2872\\
Total (GPU)&3.5485\\
\hline
\end{tabular}
\label{tab:timeeval}
\end{table}

\subsection{Comparison on the Yale's RBC dataset}
Since each of researchers usually has their own datasets which are different in the number of classes and the number of samples, thus the method comparison is quite not straightforward.  However,  we found an available RBC dataset used in Durant et.al. \cite{durant_very_2017} provided by the Yale University School of Medicine.  Their dataset contains 3,737 labeled RBCs with 10 classes including the overlapping cells. Durant et.al. used DenseNet \cite{huang_densely_2018} which has more than 150 layers. The reported accuracy was 0.9692 on the test set.

\begin{figure}[]
    \centering
    \subfigure[\newline Prediected:Dacrocyte\newline True class:Schistocyte]
    {\includegraphics[width=3.6cm, height=3.6cm]{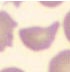}}
    \subfigure[\newline Prediected:Schistocyte\newline True class:Dacrocyte]
    {\includegraphics[width=3.6cm, height=3.6cm]{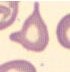}}
    \subfigure[\newline Prediected:Schistocyte\newline True class:Echinocyte ]
    {\includegraphics[width=3.6cm, height=3.6cm]{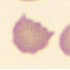}}
    \subfigure[\newline Prediected:Acanthocyte\newline True class:Schistocyte]
    {\includegraphics[width=3.6cm, height=3.6cm]{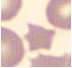}}
    \subfigure[\newline Prediected:Dacrocyte\newline True class:Elliptocyte]
    {\includegraphics[width=3.6cm, height=3.6cm]{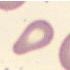}}
    \subfigure[\newline Prediected:Spherocyte\newline True class:Normal]
    {\includegraphics[width=3.6cm, height=3.6cm]{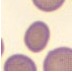}}

    \caption{Among the 748 test images tested on EfficientNet-B1 using Yale's dataset, there were six misclassified images (a)-(f)}
    \label{fig:rbcmisslabel}
\end{figure}

To make a fair comparison, we employed our proposed method based on the EfficientNet-B1 without the overlapping cell separation because the dataset contains overlapping cells as a class. The DenseNet-169 that was used in \cite{durant_very_2017} was retrained on our environment. We also used five-fold cross validation for training because we do not know how the data was partitioned in the \cite{durant_very_2017}.  Our result yields 0.9813 on the average accuracy on cross validation, and the highest and lowest cross validation accuracies are 0.9920 and 0.9733 respectively.  Table \ref{tab:precisionpublicdataset} shows our average precision, recall, and F1-Score of five-fold cross validation using EfficientNet-b1 and DenseNet on Yale’s dataset. Table \ref{tab:confusionmatrixb1} shows confusion matrix for our classifier based on EfficientNet-B1. There were only 6 wrong predicted results, as shown in Figure \ref{fig:rbcmisslabel}. 
According to the comparison result on the same dataset, our proposed method outperforms the previous work done in \cite{durant_very_2017} by yielding the higher accuracy. The overlapping cells were all correctly predicted in all five-fold cross validation which is quite obvious because the area of an overlapping cell is typically larger than other types of a single cell.  Although accuracy gain using our model compared with the previous method is about 0.18\% (7/3,737) which is not quite significant, but our model yield also better performance on both training and inference due to lots lower number of parameters.

\begin{table}[]
\caption{Average Precision, Recall, and F1-score of five-fold cross validation using EfficientNet-b1 and DenseNet-169 on Yale's dataset}
\label{tab:precisionpublicdataset}
\centering
\begin{tabular}{l|c|c|c|c|c|c|}
\cline{2-7}
&\multicolumn{3}{c|}{EfficientNet-b1} & \multicolumn{3}{c|}{DenseNet}\\
\hline
 \multicolumn{1}{|c|}{RBC Types} & Precision & Recall & F1-score & Precision & Recall & F1-score \\ \hline
\multicolumn{1}{|c|}{Normal} & 0.9885 & 0.9951 & 0.9917 & 0.9893 & 0.9951 & 0.9921\\ 
\multicolumn{1}{|c|}{Echinocyte} & 0.9907 & 0.9937 & 0.9920 & 0.9875 & 0.9968 & 0.9921\\ 
\multicolumn{1}{|c|}{Dacrocyte} & 0.8552 & 0.9000 & 0.8693 & 0.8173 & 0.9000 & 0.8535\\ 
\multicolumn{1}{|c|}{Schistocyte} & 0.9799 & 0.9652 & 0.9700 & 0.9760 & 0.9553 & 0.9653\\ 
\multicolumn{1}{|c|}{Elliptocyte} & 0.9293 & 0.9889 & 0.9576 & 0.9584 & 0.9889 & 0.9730\\ 
\multicolumn{1}{|c|}{Acanthocyte} & 0.9572 & 0.9394 & 0.9481 & 0.9627 & 0.9272 & 0.9445\\ 
\multicolumn{1}{|c|}{Target cell} & 0.9972 & 0.9972 & 0.9972 & 0.9972 & 0.9945 & 0.9959\\ 
\multicolumn{1}{|c|}{Stomatocyte} & 0.9572 & 0.9818 & 0.9691 & 0.9489 & 0.9818 & 0.9648\\ 
\multicolumn{1}{|c|}{Spherocyte} & 0.9914 & 0.9625 & 0.9767  & 0.9711 & 0.9625 & 0.9666\\ 
\multicolumn{1}{|c|}{Overlap} & 1.0000 & 1.0000 & 1.0000 & 1.0000 & 1.0000 & 1.0000\\ \hline
\end{tabular}
\end{table}

    \begin{table*}[]
\caption{Confusion matrix for our proposed method on Yale's dataset}
\label{tab:confusionmatrixb1}
\begin{adjustbox}{width=\textwidth}
\begin{tabular}{l|c|c|c|c|c|c|c|c|c|c|}
\cline{2-11}
&\multicolumn{10}{c|}{True Class}\\
\hline
 \multicolumn{1}{|c|}{Predicted Class}& Normal & Echinocyte & Dacrocyte & Schistocyte & Elliptocyte  &Acanthocyte & Target& Stomatocyte & Spherocyte & Overlap \\ \hline
 \multicolumn{1}{|l|}{Normal} & \textbf{203} & 0 & 0 & 0 & 0 & 0 & 0 & 0 & 0 & 0 \\ \hline
\multicolumn{1}{|l|}{Echinocyte}  & 0 & \textbf{63} & 0 & 0 & 0 & 0 & 0 & 0 & 0 & 0 \\ \hline
\multicolumn{1}{|l|}{Dacrocyte} & 0 & 0 & \textbf{16} & 1 & 1 & 0 & 0 & 0 & 0 & 0 \\ \hline
\multicolumn{1}{|l|}{Schistocyte}  & 0 & 1 & 1 & \textbf{150} & 0 & 0 & 0 & 0 & 0 & 0 \\ \hline
\multicolumn{1}{|l|}{Elliptocyte} & 0 & 0 & 0 & 0 & \textbf{18} & 0 & 0 & 0 & 0 & 0 \\ \hline
\multicolumn{1}{|l|}{Acanthocyte} & 0 & 0 & 0 & 1 & 0 & \textbf{32} & 0 & 0 & 0 & 0 \\ \hline
\multicolumn{1}{|l|}{Target cell}  & 0 & 0 & 0 & 0 & 0 & 0 & \textbf{145} & 0 & 0 & 0 \\ \hline
\multicolumn{1}{|l|}{Stomatocyte} & 0 & 0 & 0 & 0 & 0 & 0 & 0 & \textbf{22} & 0 & 0 \\ \hline
\multicolumn{1}{|l|}{Spherocyte} & 1 & 0 & 0 & 0 & 0 & 0 & 0 & 0 & \textbf{47} & 0 \\ \hline
\multicolumn{1}{|l|}{Overlap} & 0 & 0 & 0 & 0 & 0 & 0 & 0 & 0 & 0 & \textbf{46} \\ \hline
\end{tabular}
\end{adjustbox}
\end{table*}

\section{Conclusion \& future works}
Herein, a method to segment RBCs is presented that has the ability to separate overlapping cells based on concave points, and classify RBCs into 12 classes. The process started from color normalization, which reduced the color variations and allowed the trained model to not be biased on color. Next, contour extraction was used to extract the RBC contour from the background. Then, overlapping cells were separated using a new method to find concave points and use direct ellipse fitting on the set of detected convex curves to estimate the shape of a single RBC. Lastly, classification using EfficientNet-B1 showed the best result with augmentation. Moreover, further analysis for handling an imbalanced dataset revealed that weight balancing has the ability to reduce the bias of a trained model on the majority classes.

Many deep learning studies on RBCs still lack a standard public dataset to evaluate their performance. Our dataset has more samples and more types of RBCs than many previous studies, but it still requires to be improved for imbalanced problems. We make this dataset publically available to share other researchers and hope that it will accelerate new innovation for RBC blood smear detection. For the method presented here, we used the EfficientNet model to classify the RBCs. However, the segmentation step is not a learning-based method, which is a trend that has shown better results in many specific computer vision areas. Future work will evaluate the use of object detection methods to find a bounding box and classify the RBCs.


\bibliographystyle{elsarticle-num} 
\bibliography{reference}





\end{document}